\documentclass[
preprint,
floatfix,
aps,
prl,
superscriptaddress,
amssymb,
]{revtex4}

\usepackage{float}
\usepackage{amssymb}
\usepackage{latexsym}
\usepackage[pdftex]{graphicx}
\usepackage{amsmath}
\usepackage{graphicx}
\usepackage{dcolumn}
\usepackage{amsfonts}
\usepackage{bm}
\usepackage{epsfig}

\newcommand{\be}{\begin{equation}}
\newcommand{\ee}{\end{equation}}
\newcommand{\bea}{\begin{eqnarray}}
\newcommand{\eea}{\end{eqnarray}}

\newcommand\fpk{\mbox{$f_{\mathrm{pk}}$}}
\def\sf{S/f_{\mathrm{pk}}}
\def\sfe{(S/f_{\mathrm{pk}})^{\mathrm{exp}}}
\def\sft{(S/f_{\mathrm{pk}})^{\mathrm{sr}}}

\def\ns{\mbox{$\nu^{\star}$}}

\def\to{t_{\mathrm{0}}}
\def\zo{Z_{\mathrm{0}}}
\def\lb{\ell_{\mathrm{B}}}

\begin{document}
\title{Microwave spectroscopic observation of a Wigner solid within  the 1/2 fractional quantum Hall effect
}

\author{A.\,T. Hatke}
\email[Corresponding author: ]{hatke@magnet.fsu.edu}
\affiliation{National High Magnetic Field Laboratory, Tallahassee, Florida 32310, USA}

\author{Yang\,Liu}
\affiliation{Department of Electrical Engineering, Princeton University, Princeton, New Jersey 08544, USA}

\author{L.\,W. Engel}
\affiliation{National High Magnetic Field Laboratory, Tallahassee, Florida 32310, USA}

\author{L.\,N. Pfeiffer}
\affiliation{Department of Electrical Engineering, Princeton University, Princeton, New Jersey 08544, USA}

\author{K.\,W. West}
\affiliation{Department of Electrical Engineering, Princeton University, Princeton, New Jersey 08544, USA}

\author{K.\,W. Baldwin}
\affiliation{Department of Electrical Engineering, Princeton University, Princeton, New Jersey 08544, USA}
 
\author{M. Shayegan}
\affiliation{Department of Electrical Engineering, Princeton University, Princeton, New Jersey 08544, USA}
\received{\today}

\maketitle

\textbf{The fractional quantum Hall effect (FQHE) states at half integer Landau fillings ($\nu$) have long been of great interest  \cite{suen:1992,suen:1994,shabani:2013,liu:2014a,liu:2014b,falson:2015,willett:2013,eisenstein:1992}, since they have correlations that differ from those of the fundamental Laughlin states found at odd denominators.
At $\nu=1/2$ the FQHE has been observed in wide \cite{suen:1992,suen:1994,shabani:2013,liu:2014a,liu:2014b} or double quantum wells \cite{eisenstein:1992}, and is ascribed to the two-component Halperin-Laughlin $\Psi_{331}$ state \cite{halperin:1983,multicomprvw}.
$\Psi_{331}$ excitations carry charge $\pm e/4$, like the carriers of $\nu=5/2$ states which are of interest in quantum computation \cite{willett:2013}. 
Further, such an excitation (quasiparticle or -hole) of  $\Psi_{331}$ has unequal, opposite charge in the top and bottom layers, and hence an up or down dipole moment  \cite{multicomprvw} .       
Here we report evidence for a Wigner solid (WS) of such dipolar quasiholes (see Fig. 1a) from a quantitative study of the microwave spectra of a wide quantum well (WQW) at $\nu$ close to 1/2.
}

WSs in quantum Hall systems can be classified into two types \citep{archer:2011}.
Here we take a type-I WS to be a ground state of the entire system, such as is found at the low $\nu$ termination of the FQHE series \citep{lozovik:1975,andrei:1988,williams:1991,kukushkin:1994,paalanen:1992b,yewc,msreview}.
Type-II WSs are formed of quasiparticles or quasiholes in the presence of a gapped state such as a filled Landau level at an integer quantum Hall effect (IQHE) plateau \citep{chen:2003,hatke:2014,tiemann:2014}, or an FQHE liquid \citep{zhu:2010}.  
The dipole solid that we report here is of type II. 
Crucially, the density of carriers in a type-II WS, but not a type-I WS, is dependent on the magnetic field, because the carrier density increases as flux is subtracted or added to the parent, gapped state.

In a WQW the $1/2$ FQHE state exists for electron density, $n$, within a certain range \citep{suen:1992,suen:1994,manoharan:1996,shabani:2013}.  
Experiments showed that increasing $n$ in a WQW  causes its carriers  to become more bilayer-like by reducing the first-excited subband energy and by increasing the interlayer distance and the intralayer Coulomb interaction, $e^2/4\pi\varepsilon\lb$, where $\lb=\sqrt{\hbar/eB}$ is the magnetic length.
Transport studies further showed \citep{suen:1992,suen:1994,manoharan:1996,shabani:2013} that the  $1/2$ FQHE state occurs when the system is sufficiently bilayer-like, but only when the inter- and intralayer Coulomb interaction energies are still comparable.
As $\nu$ in a WQW is decreased from 1/2 by increasing $B$, the FQHE gives way to an insulator \citep{manoharan:1996}, which extends to the lowest $\nu$ measured.  
That insulator is interpreted as a type-I WS pinned by disorder, and a recent theory \citep{goerbigwqw} of WQWs identifies this WS as a pinned bilayer rectangular lattice of the first excited subband.

The microwave spectroscopic measurements reported  here are of WS pinning modes. 
The pinning mode is a well-known experimental signature of WS \citep{andrei:1988,williams:1991,paalanen:1992b,yewc}, and corresponds to an oscillation of the solid within the potential of the disorder that pins it. 
For $\nu$ just below $1/2$ we  find a resonance that can be identified as the pinning mode of the type-II 1/2 FQHE quasihole WS.  
The resonance intensity allows measurement of the participating carrier density by means of a sum rule. 
For the resonance interpreted as  type II, we find  that participating density  measured this way depends linearly on $1/2-\nu$, quantitatively just as expected for quasiholes of the 1/2 FQHE.  
This resonance also is suppressed by a small imbalance between the carrier density in top and bottom layers, consistent with its interpretation as stemming from a solid of quasiholes of the 1/2 FQHE ground state. 
Along with the pinning mode of the type-II 1/2 FQHE quasihole WS, we also find a resonance due to the type-I WS that extends to lowest $\nu$.  
The amplitude of the  pinning mode of the type-I WS increases monotonically as $\nu$ decreases below 1/2, and the presence of two resonances in the spectra indicates the system is  a mixture of two WS phases.

\begin{figure*}[t]
\vspace{-0.2 in}
\includegraphics{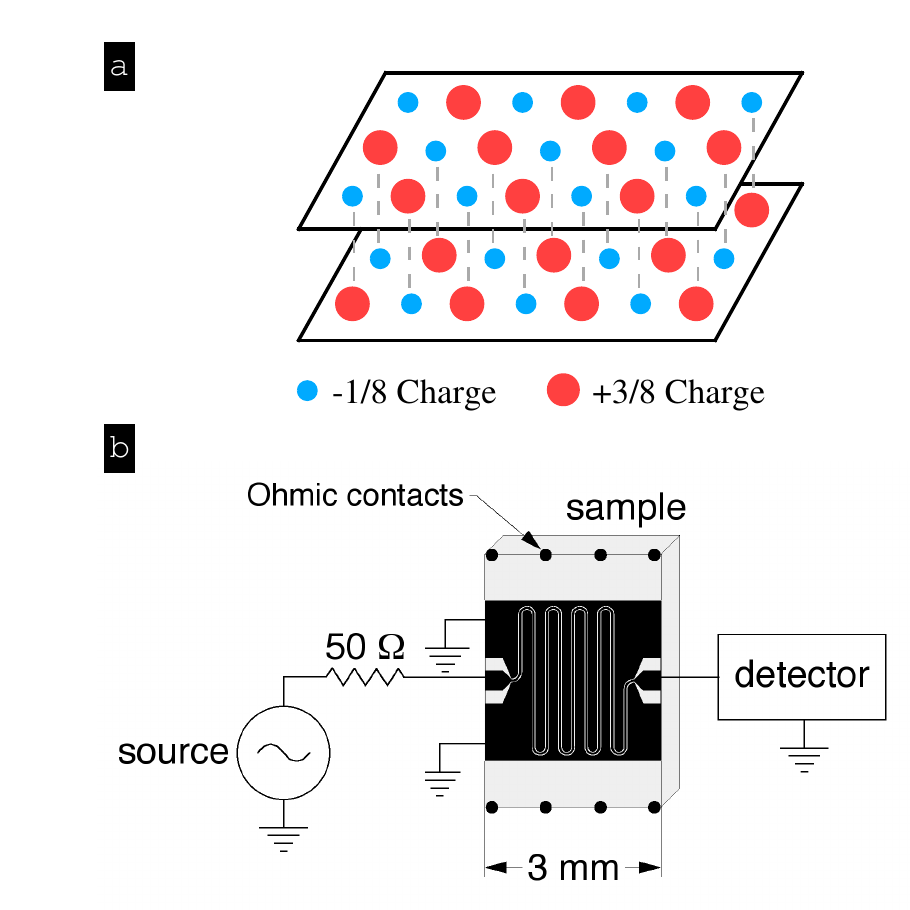}
\vspace{-0.2 in}
\caption{\textbf{$\nu=1/2$ FQHE quasihole solid and microwave setup.} 
(a) Sketch of a possible structure for a dipolar WS of quasiholes of the 1/2 FQHE.   
The quasiholes each have charge $+3e/8$ (shown as red)  in one layer, and $-e/8$ (shown as blue)  in the other. 
(b) Schematic of the microwave measurement set-up.
The source and detector are outside the cryostat at room temperature and the coplanar waveguide transmission line is patterned in metal film on top of the sample surface.   
Metal film of the transmission line is shown as black. 
The microwave conductivity $\sigma_{xx}$ is calculated from loss through the line. 
}
\label{cartoon}
\end{figure*}

Measurements were performed on a GaAs/AlGaAs WQW of width $w=80\,$nm with an as-cooled density of $n=1.1$ in units of $10^{11}\,$cm$^{-2}$, which we use for density throughout the paper.
Figure\,\ref{cartoon}\,(b) shows a top view schematic of our coplanar waveguide technique \citep{yewc,msreview,chen:2003,chen:2004}.
We maintained a symmetric growth-direction charge distribution about the well center  unless otherwise noted (see Methods section).    
\begin{figure}[t]
\vspace{-0.2 in}
\includegraphics[width=.5\textwidth]{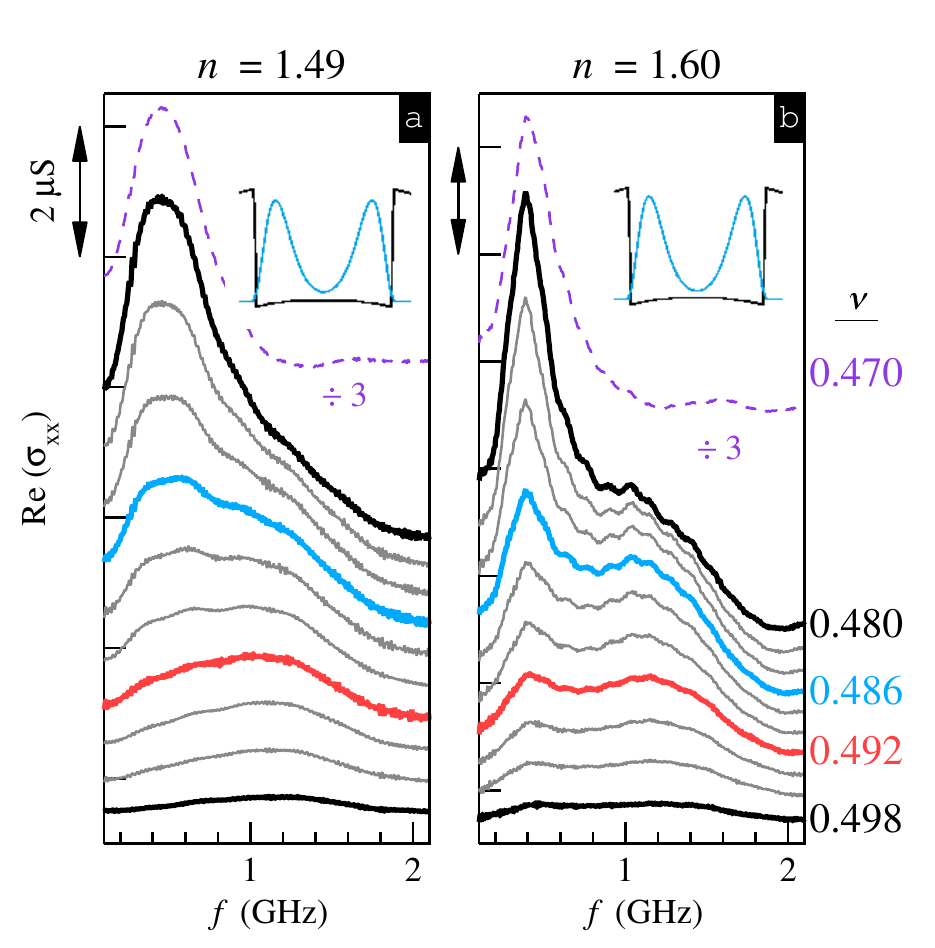}
\vspace{-0.2 in}
\caption{\textbf{Microwave spectra near $\nu=1/2$.} 
(a) and (b) Microwave spectra, plotted as the real part of the conductivity (Re\,$(\sigma_{xx})$) vs the frequency $(f)$ for $n=1.49$ and $n=1.60$ in units of $10^{11}\,$cm$^{-2}$.
The $\nu=0.470$ trace is shown as a dashed line with values divided by 3; the weak bump at $f=1.6\,$GHz is an artifact (see Methods section). 
Traces from $\nu=0.480$ to $\nu=0.498$ are shown with $0.002$ steps, and are successively offset by $0.5\,\mu$S.
Insets: Simulation of the growth-direction charge distribution at the specified $n$,  when the charge is symmetric about the well center.
}
\label{spectra}
\end{figure}

In Fig.\,\ref{spectra}\,(a) we plot the real part of the conductivity, Re\,$(\sigma_{xx})$, vs frequency, $f$, for several $\nu$, at $n=1.49$.
At $\nu=0.498$ we observe a resonance with peak frequency $\fpk \sim 1.2\,$GHz.
As $\nu$ decreases this resonance has nonmonotonic amplitude variation.
An additional resonance is observed at lower $\nu$ with $\fpk \sim 0.45\,$GHz.
We ascribe this lower-$\fpk$ resonance to a type-I WS, whose $\fpk$ is consistent with that found in an earlier study \citep{hatke:2015}.
In Fig.\,\ref{spectra}\,(b) at $n=1.60$ we observe two well-defined resonances throughout the $0.480\leq\nu<0.5$ range.
We interpret the higher-$\fpk$ resonance as due to the type-II WS composed of quasiholes of the $1/2$ FQHE, which coexists with the type-I bilayer electron solid.

\begin{figure}[b]
\includegraphics[width=.5\textwidth]{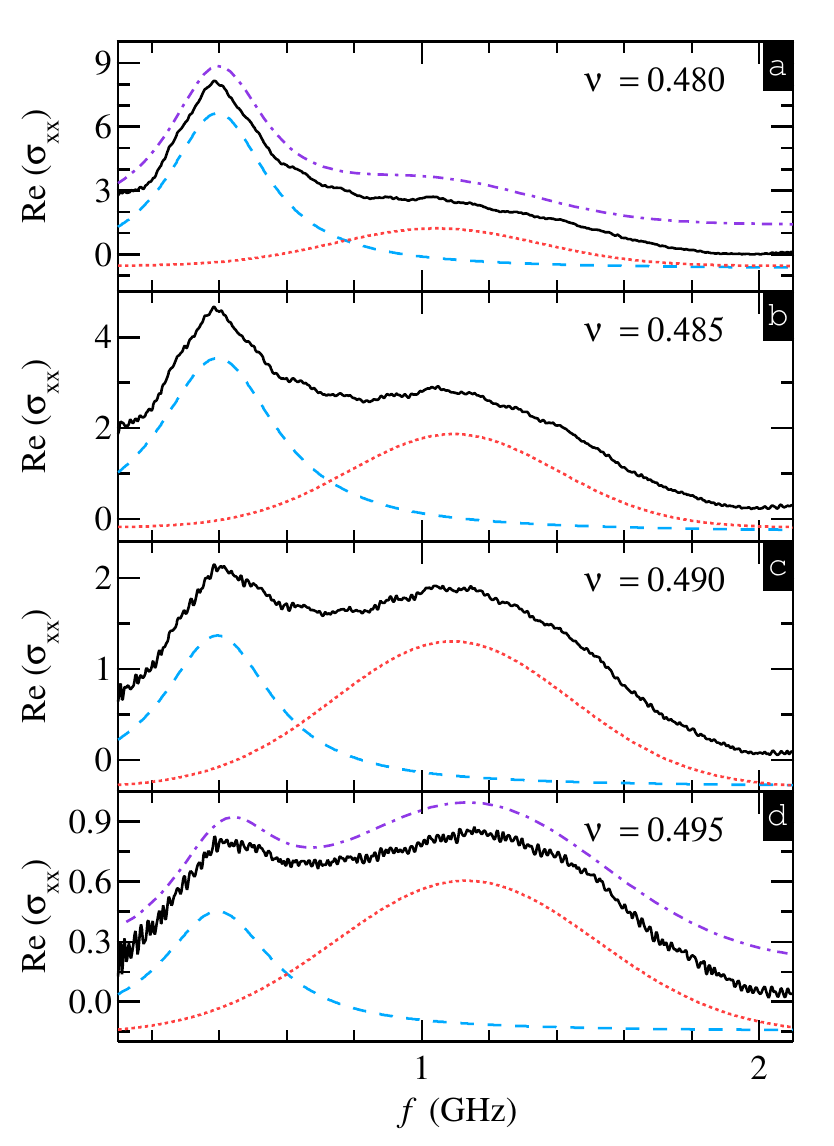}
\vspace{-0.2 in}
\caption{\textbf{Two resonance decomposition.} 
(a)-(d) Re\,$(\sigma_{xx})$ vs $f$ at several $\nu$ (solid traces) for $n=1.60$.
Each plot also contains the components of the individual peak fits for the type-I solid resonance (dashed line) and the $\nu=1/2$ FQHE quasihole solid (dotted line), which are vertically offset for clarity.
$\fpk$ for the type-I WS resonance is held constant and $\fpk$ for the type-II WS resonance is allowed to vary for all fits.
In (a) and (d) the total fitting function is shown as a dashed-dotted line, which is also vertically offset for clarity.
}
\label{coex}
\end{figure}
To test our interpretation we compare the charge density obtained from the intensity of the type-II resonance to that expected for a type-II WS of quasiholes. 
We reproduce four spectra from Fig.\,\ref{spectra}\,(b) in Figs.\,\ref{coex}\,(a)-(d).
The spectra are fitted to a sum of two single-peak functions. 
These fits are used to extract the two resonances over the entire $\nu$ range studied for $n=1.60$ and $n=1.49$, and to calculate the integrated intensities, $S=\,\int$Re\,$[\sigma_{xx}(f)] df$.

\begin{figure}[t]
\includegraphics[width=.49\textwidth]{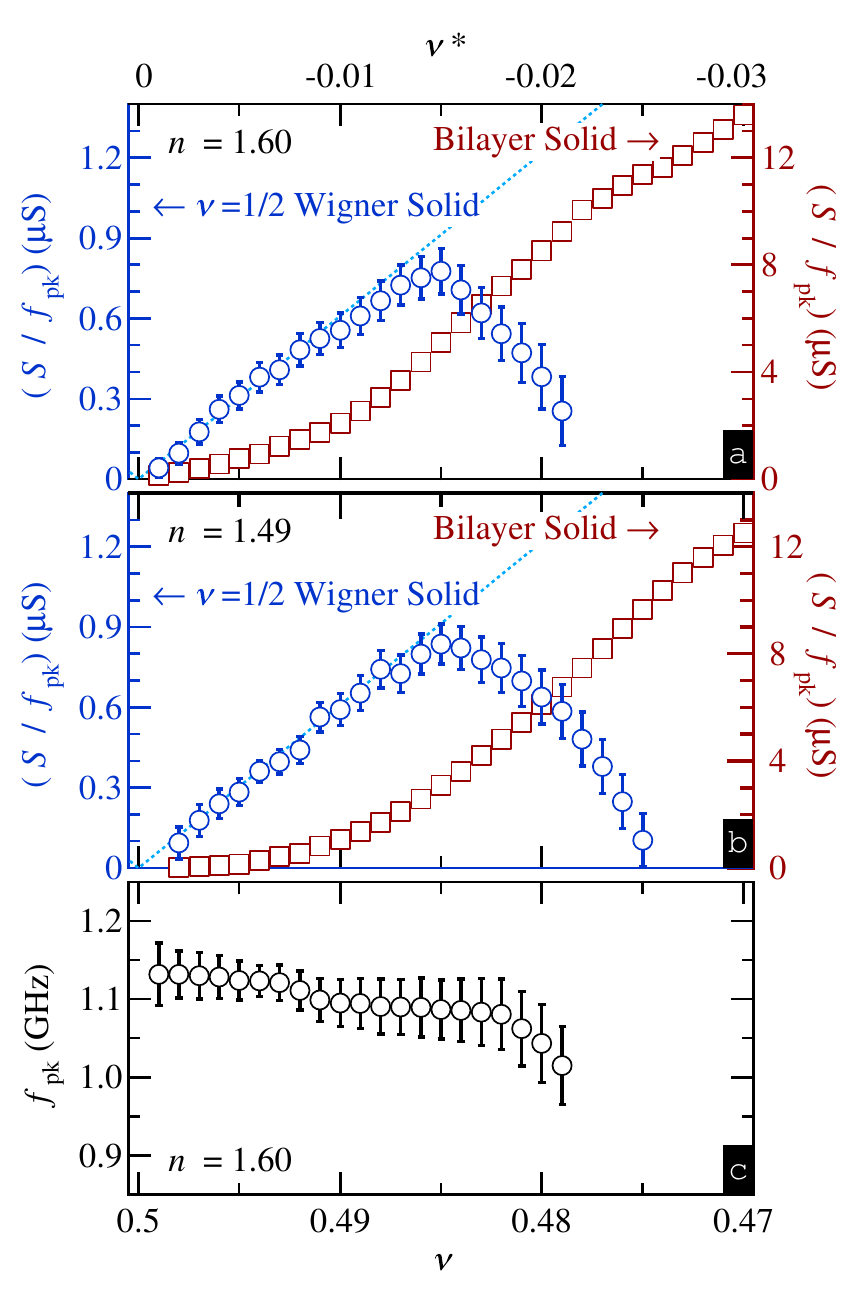}
\vspace{-0.2 in}
\caption{\textbf{Pinning mode analysis.} 
(a) and (b) Integrated spectrum divided by peak frequency, $\sfe$, vs $\nu$ for the resonance of  type-II WS (open circles, left axis), and the type-I WS (open squares, right axis). 
The scales differ by a factor of $10$. 
The dotted line in each plot corresponds to the sum rule $\sft$ \citep{fukuyama:1978} for full participation of quasiholes, and refers to the left axes.  
For (a) $n=1.60$ and (b) $n=1.49$. 
(c) $\fpk$ for the type-II WS vs $\nu$ for $n=1.60$; on the top axis  $\nu^{\star}=1/2-\nu$ is shown.
Error bars are estimated from using different fit functions and ranges.}
\label{sfpk}
\end{figure}

Pinning modes roughly obey a sum rule \citep{fukuyama:1978} $\sft=\rho_c \pi/2B=e^{2}\pi \tilde{\nu} /2h$, where $\tilde{\nu}=\nu$ for a type-I WS, or $\tilde{\nu}=\nu^{\star}=\nu-1/2$ for a type-II WS of the 1/2 FQHE. 
The charge density, $\rho_c$, of the relevant carriers is $ne|\nu^\star|/\nu$ for the type-II WS near $\nu=1/2$ and $ne$ for the type-I WS.
In Figs.\,\ref{sfpk} (a) and (b) we plot the experimentally obtained $\sf$, $\sfe$, for the two resonances (symbols) and the calculated $\sft$ (dotted line).  
$\sfe$ of the type-II WS is in good agreement with $\sft$, with no adjustable  parameters, for $0.485<\nu<0.5$.
For $\nu<0.485$ $\sfe$ of the type-II WS resonance deviates from $\sft$ and begins to decrease with increasing quasihole density.
The good agreement between $\sfe$ and $\sft$ provides compelling evidence for the interpretation of the higher-$\fpk$ resonance as due to a type-II WS made up of quasiholes.

The close agreement for the type-II WS of $\sfe$ and $\sft$ implies that only a small percentage of  carriers are available for the type-I WS.
We investigated the participation ratio $\eta\equiv \sfe/\sft$ for the type-I WS.  
At $\nu=0.485$, at which $\sfe$ begins to fall short of $\sft$ for the type-II WS, we find $\eta\sim 10\%$ for $n=1.49$ and $\eta\sim 17\%$ for $n=1.60$.
Since $\eta$ of the type-I WS resonance continues to decrease as $\nu$ goes from $0.485$ toward $1/2$, near-full participation of the carriers in the type-II WS resonance is consistent with the observed type-I WS resonance intensity. 
When $\nu$ goes below $0.48$ the type-I WS resonance amplitude increases as the electrons solidify into the type-I WS; by $\nu=0.4$, $\eta\sim 100\%$ for the type-I WS \citep{hatke:2015}.

Figure\,\ref{sfpk}\,(c) shows the extracted $\fpk$ vs $\nu$ for the type-II WS, for $n=1.60$ (data fro $n=1.49$  yield nearly identical values).
$\fpk$ decreases by $\sim 5\%$ for $0.5>\nu>0.485$ with larger decreases for $\nu<0.485$.
One would expect \citep{chitra:2001,fertig:1999,fogler:2000} that $\fpk$ should decrease as $\nu^{\star}$ (and hence the quasihole density) increases due to the increase of the WS shear modulus.
However, the local density within domains of the type-II solid can remain constant while  the total proportion of the area occupied by these domains increases proportional to $\ns$.     
The overall picture of the system near $\nu=1/2$ is then of a composite, with 1) domains of local $\nu=1/2$ FQHE, 2) domains of type-II WS of quasiholes of that state, and 3) domains of a type-I WS.

\begin{figure}[t]
\includegraphics[width=.49\textwidth]{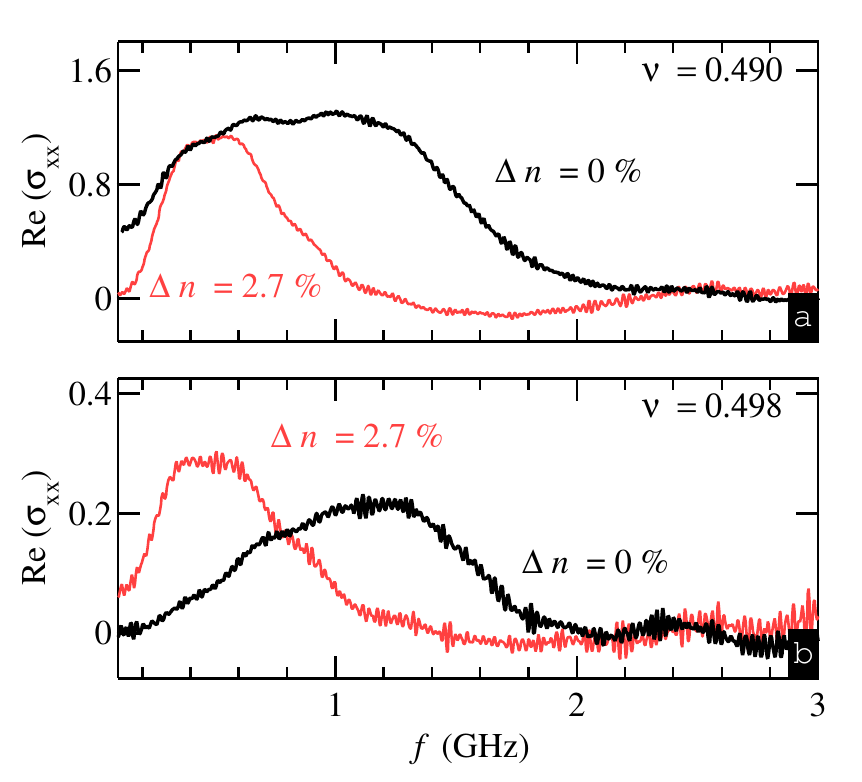}
\vspace{-0.2 in}
\caption{\textbf{Microwave spectra with asymmetric gating.} 
Re\,$(\sigma_{xx})$ vs $f$ at fixed $\nu$ for (a) $\nu=0.490$ and (c) $\nu=0.498$.
Spectra are obtained at fixed total density, $n=1.49$, and different charge configurations along the growth direction within the well, with $\Delta n$ denoting the charge difference between the top and bottom layers, specified as a percentage of the total charge density $n$. 
}
\label{asym}
\end{figure}
In WQWs the $1/2$ FQHE state has been shown to be highly sensitive to the symmetry of the growth-direction charge distribution  \citep{suen:1994,manoharan:1996,shabani:2013}.
To further test the association of the higher-$\fpk$ resonance with the 1/2 FQHE we investigated the role of charge asymmetry between the two layers (see Methods section).

Figure\,\ref{asym} shows Re\,$(\sigma_{xx})$ vs $f$ at $n=1.49$ for two fixed $\nu$.
In Fig.\,\ref{asym}\,(a) ($\nu=0.490$) the trace for the symmetric case contains both resonances. 
We observe a loss of the higher-$\fpk$, type-II WS resonance, and a concurrent enhancement of the type-I WS resonance near $f\simeq 0.45\,$GHz on imbalancing the layer densities by $\Delta n=n_f-n_b/n=2.7\%$, where $n_f$ and $n_b$ are respectively the front and back layer densities in the WQW.  
This result is more dramatic at $\nu=0.498$, shown in Fig.\,\ref{asym}\,(b), where the symmetric state shows \textit{only} the higher-$\fpk$ type-II WS resonance  while the $\Delta n=2.7 \%$ case shows \textit{only} the type-I resonance. 
The loss of the type-II WS with this  slight asymmetry of the charge distribution is explained if the $1/2$ FQHE and its excitations are much more sensitive to the symmetry than the type-I WS.  
These $\Delta n$ values are in reasonable agreement with Ref.\,\cite{ manoharan:1996}.
Larger asymmetries of our WQW ($|\Delta n|\geq10\%$) destroy the type-I WS resonance as well \citep{manoharan:1996,hatke:2015}.

In dc transport measurements  a reentrant insulating range was observed at $\nu$ slightly larger than $\nu=1/2$ \citep{manoharan:1996}.
Our microwave measurements for $\nu>1/2$ displayed no resonance; there was no observable pinning mode of the quasiparticles associated with $\nu=1/2$ FQHE. 
A plausible explanation is that the quasiholes and quasiparticles have different interactions. 
Such a situation was predicted theoretically \citep{archer:2011} near $\nu=1/3$ for which the quasihole solid was calculated to have a higher melting temperature than the quasiparticle solid.

In summary, our study of a WQW provides strong evidence of a type-II WS of 1/2 FQHE quasiholes, which have unequal opposite-sign charges in the two layers.  
The solid appears to coexist along with the type-I solid characteristic of the WQW system at low $\nu$.

\section{Methods}
\subsection{Microwave spectroscopy}
Our microwave spectroscopy technique \cite{chen:2004,hatke:2014} uses a coplanar waveguide (CPW) on the surface of a sample.
A NiCr front gate was deposited on glass that was etched to space it from the CPW by $\sim 10\,\mu$m.
A schematic diagram of the microwave measurement technique is shown in Fig.\,\ref{cartoon}\,(b).
We calculate the diagonal conductivity as $ \sigma_{xx} (f) = (s/ l \zo) \ln (t/\to)$, where $s=30\ \mu$m is the distance between the center conductor and ground plane, $l=28\,$mm is the length of the CPW, $\zo=50\,\Omega$ is the characteristic impedance without the 2DES, $t$ is the transmitted signal amplitude and $\to$ is the normalizing amplitude taken at $\nu=1/2$.
The microwave measurements were carried out in the low-power limit, in which the measurement is not sensitive to the excitation power, at a bath temperature of $T=50\,$mK.

Connection of the sample to the mount resulted in a weak bump near $\sim 1.6\,$GHz, Fig.\,\ref{spectra}\,(a) and (b) that is an artifact: the strong line at $\nu=0.470$ has significant Im$(\sigma_{xx})$ at 1.6 GHz, which enables a standing wave between the line and a reflection near the sample mounting. 
$|$Im$(\sigma_{xx})|$ is much smaller for the higher $\nu$ data.

\subsection{Charge distribution}
As described elsewhere \citep{suen:1994,manoharan:1996,shabani:2013,liu:2014a,hatke:2014,hatke:2015}, bias voltages applied to back and front gates control the carrier density and the symmetry of the growth-direction charge distribution.
The back gate was in direct contact with the bottom of the sample and the front gate was deposited on a piece of glass that was etched to be spaced from the sample surface to not interfere with the microwave transmission line. 
A symmetric (balanced) growth-direction charge distribution was maintained by biasing the gates such that individually they would change the density by the same amount with equal and opposite electric fields. 
An asymmetric (imbalanced) distribution at a fixed density was obtained by first biasing one gate to get half the desired charge asymmetry, $\Delta n / 2$. 
Then we biased the other gate with opposite polarity to recover the original total density.

\section{Acknowledgements}
We thank N. Bonesteel, K. Yang, and K. Muraki for discussions.
We also thank Ju-Hyun Park and Glover Jones for technical assistance.
The microwave spectroscopy work at NHMFL was supported through DOE grant DE-FG02-05-ER46212 at NHMFL/FSU.   
The National High Magnetic Field Laboratory (NHMFL), is supported by NSF Cooperative Agreement No. DMR-0654118, by the State of Florida, and by the DOE.   
The work at Princeton University was funded by the Gordon and Betty Moore Foundation through the EPiQS initiative Grant GBMF4420, and by the National Science Foundation through grant DMR-1305691 and MRSEC Grant DMR-1420541.

Data displayed in this manuscript will be available by email request to engel@magnet.fsu.edu.

\bibliographystyle{apsrev}

\begin{thebibliography}{30}
\expandafter\ifx\csname natexlab\endcsname\relax\def\natexlab#1{#1}\fi
\expandafter\ifx\csname bibnamefont\endcsname\relax
  \def\bibnamefont#1{#1}\fi
\expandafter\ifx\csname bibfnamefont\endcsname\relax
  \def\bibfnamefont#1{#1}\fi
\expandafter\ifx\csname citenamefont\endcsname\relax
  \def\citenamefont#1{#1}\fi
\expandafter\ifx\csname url\endcsname\relax
  \def\url#1{\texttt{#1}}\fi
\expandafter\ifx\csname urlprefix\endcsname\relax\def\urlprefix{URL }\fi
\providecommand{\bibinfo}[2]{#2}
\providecommand{\eprint}[2][]{\url{#2}}

\bibitem[{\citenamefont{Suen et~al.}(1992)\citenamefont{Suen, Engel, Santos,
  Shayegan, and Tsui}}]{suen:1992}
\bibinfo{author}{\bibfnamefont{Y.~W.} \bibnamefont{Suen}},
  \bibinfo{author}{\bibfnamefont{L.~W.} \bibnamefont{Engel}},
  \bibinfo{author}{\bibfnamefont{M.~B.} \bibnamefont{Santos}},
  \bibinfo{author}{\bibfnamefont{M.}~\bibnamefont{Shayegan}}, \bibnamefont{and}
  \bibinfo{author}{\bibfnamefont{D.~C.} \bibnamefont{Tsui}},
  \bibinfo{journal}{Phys. Rev. Lett.} \textbf{\bibinfo{volume}{68}},
  \bibinfo{pages}{1379} (\bibinfo{year}{1992}).

\bibitem[{\citenamefont{Suen et~al.}(1994)\citenamefont{Suen, Manoharan, Ying,
  Santos, and Shayegan}}]{suen:1994}
\bibinfo{author}{\bibfnamefont{Y.~W.} \bibnamefont{Suen}},
  \bibinfo{author}{\bibfnamefont{H.~C.} \bibnamefont{Manoharan}},
  \bibinfo{author}{\bibfnamefont{X.}~\bibnamefont{Ying}},
  \bibinfo{author}{\bibfnamefont{M.~B.} \bibnamefont{Santos}},
  \bibnamefont{and} \bibinfo{author}{\bibfnamefont{M.}~\bibnamefont{Shayegan}},
  \bibinfo{journal}{Phys. Rev. Lett.} \textbf{\bibinfo{volume}{72}},
  \bibinfo{pages}{3405} (\bibinfo{year}{1994}).

\bibitem[{\citenamefont{Shabani et~al.}(2013)\citenamefont{Shabani, Liu,
  Shayegan, Pfeiffer, West, and Baldwin}}]{shabani:2013}
\bibinfo{author}{\bibfnamefont{J.}~\bibnamefont{Shabani}},
  \bibinfo{author}{\bibfnamefont{Y.}~\bibnamefont{Liu}},
  \bibinfo{author}{\bibfnamefont{M.}~\bibnamefont{Shayegan}},
  \bibinfo{author}{\bibfnamefont{L.~N.} \bibnamefont{Pfeiffer}},
  \bibinfo{author}{\bibfnamefont{K.~W.} \bibnamefont{West}}, \bibnamefont{and}
  \bibinfo{author}{\bibfnamefont{K.~W.} \bibnamefont{Baldwin}},
  \bibinfo{journal}{Phys. Rev. B} \textbf{\bibinfo{volume}{88}},
  \bibinfo{pages}{245413} (\bibinfo{year}{2013}).

\bibitem[{\citenamefont{Liu et~al.}(2014{\natexlab{a}})\citenamefont{Liu,
  Graninger, Hasdemir, Shayegan, Pfeiffer, West, Baldwin, and
  Winkler}}]{liu:2014a}
\bibinfo{author}{\bibfnamefont{Y.}~\bibnamefont{Liu}},
  \bibinfo{author}{\bibfnamefont{A.~L.} \bibnamefont{Graninger}},
  \bibinfo{author}{\bibfnamefont{S.}~\bibnamefont{Hasdemir}},
  \bibinfo{author}{\bibfnamefont{M.}~\bibnamefont{Shayegan}},
  \bibinfo{author}{\bibfnamefont{L.~N.} \bibnamefont{Pfeiffer}},
  \bibinfo{author}{\bibfnamefont{K.~W.} \bibnamefont{West}},
  \bibinfo{author}{\bibfnamefont{K.~W.} \bibnamefont{Baldwin}},
  \bibnamefont{and} \bibinfo{author}{\bibfnamefont{R.}~\bibnamefont{Winkler}},
  \bibinfo{journal}{Phys. Rev. Lett.} \textbf{\bibinfo{volume}{112}},
  \bibinfo{pages}{046804} (\bibinfo{year}{2014}{\natexlab{a}}).

\bibitem[{\citenamefont{Liu et~al.}(2014{\natexlab{b}})\citenamefont{Liu,
  Hasdemir, Kamburov, Graninger, Shayegan, Pfeiffer, West, Baldwin, and
  Winkler}}]{liu:2014b}
\bibinfo{author}{\bibfnamefont{Y.}~\bibnamefont{Liu}},
  \bibinfo{author}{\bibfnamefont{S.}~\bibnamefont{Hasdemir}},
  \bibinfo{author}{\bibfnamefont{D.}~\bibnamefont{Kamburov}},
  \bibinfo{author}{\bibfnamefont{A.~L.} \bibnamefont{Graninger}},
  \bibinfo{author}{\bibfnamefont{M.}~\bibnamefont{Shayegan}},
  \bibinfo{author}{\bibfnamefont{L.~N.} \bibnamefont{Pfeiffer}},
  \bibinfo{author}{\bibfnamefont{K.~W.} \bibnamefont{West}},
  \bibinfo{author}{\bibfnamefont{K.~W.} \bibnamefont{Baldwin}},
  \bibnamefont{and} \bibinfo{author}{\bibfnamefont{R.}~\bibnamefont{Winkler}},
  \bibinfo{journal}{Phys. Rev. B} \textbf{\bibinfo{volume}{89}},
  \bibinfo{pages}{165313} (\bibinfo{year}{2014}{\natexlab{b}}).

\bibitem[{\citenamefont{Falson et~al.}(2015)\citenamefont{Falson, Maryenko,
  Friess, Zhang, Kozuka, Tsukazaki, Smet, and Kawasaki}}]{falson:2015}
\bibinfo{author}{\bibfnamefont{J.}~\bibnamefont{Falson}},
  \bibinfo{author}{\bibfnamefont{D.}~\bibnamefont{Maryenko}},
  \bibinfo{author}{\bibfnamefont{B.}~\bibnamefont{Friess}},
  \bibinfo{author}{\bibfnamefont{D.}~\bibnamefont{Zhang}},
  \bibinfo{author}{\bibfnamefont{Y.}~\bibnamefont{Kozuka}},
  \bibinfo{author}{\bibfnamefont{A.}~\bibnamefont{Tsukazaki}},
  \bibinfo{author}{\bibfnamefont{J.~H.} \bibnamefont{Smet}}, \bibnamefont{and}
  \bibinfo{author}{\bibfnamefont{M.}~\bibnamefont{Kawasaki}},
  \bibinfo{journal}{Nat. Phys.} \textbf{\bibinfo{volume}{11}},
  \bibinfo{pages}{347} (\bibinfo{year}{2015}).

\bibitem[{\citenamefont{Willett}(2013)}]{willett:2013}
\bibinfo{author}{\bibfnamefont{R.~L.} \bibnamefont{Willett}},
  \bibinfo{journal}{Reports on Progress in Physics}
  \textbf{\bibinfo{volume}{76}}, \bibinfo{pages}{076501}
  (\bibinfo{year}{2013}).

\bibitem[{\citenamefont{Eisenstein et~al.}(1992)\citenamefont{Eisenstein,
  Boebinger, Pfeiffer, West, and He}}]{eisenstein:1992}
\bibinfo{author}{\bibfnamefont{J.~P.} \bibnamefont{Eisenstein}},
  \bibinfo{author}{\bibfnamefont{G.~S.} \bibnamefont{Boebinger}},
  \bibinfo{author}{\bibfnamefont{L.~N.} \bibnamefont{Pfeiffer}},
  \bibinfo{author}{\bibfnamefont{K.~W.} \bibnamefont{West}}, \bibnamefont{and}
  \bibinfo{author}{\bibfnamefont{S.}~\bibnamefont{He}}, \bibinfo{journal}{Phys.
  Rev. Lett.} \textbf{\bibinfo{volume}{68}}, \bibinfo{pages}{1383}
  (\bibinfo{year}{1992}).

\bibitem[{\citenamefont{Halperin}(1983)}]{halperin:1983}
\bibinfo{author}{\bibfnamefont{B.~I.} \bibnamefont{Halperin}},
  \bibinfo{journal}{Helv. Phys. Acta} \textbf{\bibinfo{volume}{56}},
  \bibinfo{pages}{75} (\bibinfo{year}{1983}).

\bibitem[{\citenamefont{Girvin and MacDonald}(1997)}]{multicomprvw}
\bibinfo{author}{\bibfnamefont{S.~M.} \bibnamefont{Girvin}} \bibnamefont{and}
  \bibinfo{author}{\bibfnamefont{A.~H.} \bibnamefont{MacDonald}},
  \emph{\bibinfo{title}{{\rm in }Perspectives in Quantum Hall Effects, {\rm
  edited by S. Das Sarma and A. Pinczuk}}}
  (\bibinfo{publisher}{Wiley-Interscience, New York}, \bibinfo{year}{1997}), p.
  \bibinfo{pages}{161}.

\bibitem[{\citenamefont{Archer and Jain}(2011)}]{archer:2011}
\bibinfo{author}{\bibfnamefont{A.~C.} \bibnamefont{Archer}} \bibnamefont{and}
  \bibinfo{author}{\bibfnamefont{J.~K.} \bibnamefont{Jain}},
  \bibinfo{journal}{Phys. Rev. B} \textbf{\bibinfo{volume}{84}},
  \bibinfo{pages}{115139} (\bibinfo{year}{2011}).

\bibitem[{\citenamefont{Lozovik and Yudson}(1975)}]{lozovik:1975}
\bibinfo{author}{\bibfnamefont{Y.~E.} \bibnamefont{Lozovik}} \bibnamefont{and}
  \bibinfo{author}{\bibfnamefont{V.~I.} \bibnamefont{Yudson}},
  \bibinfo{journal}{JETP Letters} \textbf{\bibinfo{volume}{22}},
  \bibinfo{pages}{11} (\bibinfo{year}{1975}).

\bibitem[{\citenamefont{Andrei et~al.}(1988)\citenamefont{Andrei, Deville,
  Glattli, Williams, Paris, and Etienne}}]{andrei:1988}
\bibinfo{author}{\bibfnamefont{E.~Y.} \bibnamefont{Andrei}},
  \bibinfo{author}{\bibfnamefont{G.}~\bibnamefont{Deville}},
  \bibinfo{author}{\bibfnamefont{D.~C.} \bibnamefont{Glattli}},
  \bibinfo{author}{\bibfnamefont{F.~I.~B.} \bibnamefont{Williams}},
  \bibinfo{author}{\bibfnamefont{E.}~\bibnamefont{Paris}}, \bibnamefont{and}
  \bibinfo{author}{\bibfnamefont{B.}~\bibnamefont{Etienne}},
  \bibinfo{journal}{Phys. Rev. Lett.} \textbf{\bibinfo{volume}{60}},
  \bibinfo{pages}{2765} (\bibinfo{year}{1988}).

\bibitem[{\citenamefont{Williams et~al.}(1991)\citenamefont{Williams, Wright,
  Clark, Andrei, Deville, Glattli, Probst, Etienne, Dorin, Foxon
  et~al.}}]{williams:1991}
\bibinfo{author}{\bibfnamefont{F.~I.~B.} \bibnamefont{Williams}},
  \bibinfo{author}{\bibfnamefont{P.~A.} \bibnamefont{Wright}},
  \bibinfo{author}{\bibfnamefont{R.~G.} \bibnamefont{Clark}},
  \bibinfo{author}{\bibfnamefont{E.~Y.} \bibnamefont{Andrei}},
  \bibinfo{author}{\bibfnamefont{G.}~\bibnamefont{Deville}},
  \bibinfo{author}{\bibfnamefont{D.~C.} \bibnamefont{Glattli}},
  \bibinfo{author}{\bibfnamefont{O.}~\bibnamefont{Probst}},
  \bibinfo{author}{\bibfnamefont{B.}~\bibnamefont{Etienne}},
  \bibinfo{author}{\bibfnamefont{C.}~\bibnamefont{Dorin}},
  \bibinfo{author}{\bibfnamefont{C.~T.} \bibnamefont{Foxon}},
  \bibnamefont{et~al.}, \bibinfo{journal}{Phys. Rev. Lett.}
  \textbf{\bibinfo{volume}{66}}, \bibinfo{pages}{3285} (\bibinfo{year}{1991}).

\bibitem[{\citenamefont{Kukushkin et~al.}(1994)\citenamefont{Kukushkin, Falko,
  Haug, von Klitzing, Eberl, and Totemayer}}]{kukushkin:1994}
\bibinfo{author}{\bibfnamefont{I.~V.} \bibnamefont{Kukushkin}},
  \bibinfo{author}{\bibfnamefont{V.~I.} \bibnamefont{Falko}},
  \bibinfo{author}{\bibfnamefont{R.~J.} \bibnamefont{Haug}},
  \bibinfo{author}{\bibfnamefont{K.}~\bibnamefont{von Klitzing}},
  \bibinfo{author}{\bibfnamefont{K.}~\bibnamefont{Eberl}}, \bibnamefont{and}
  \bibinfo{author}{\bibfnamefont{K.}~\bibnamefont{Totemayer}},
  \bibinfo{journal}{Phys. Rev. Lett.} \textbf{\bibinfo{volume}{72}},
  \bibinfo{pages}{3594} (\bibinfo{year}{1994}).

\bibitem[{\citenamefont{Paalanen et~al.}(1992)\citenamefont{Paalanen, Willett,
  Ruel, Littlewood, West, and Pfeiffer}}]{paalanen:1992b}
\bibinfo{author}{\bibfnamefont{M.~A.} \bibnamefont{Paalanen}},
  \bibinfo{author}{\bibfnamefont{R.~L.} \bibnamefont{Willett}},
  \bibinfo{author}{\bibfnamefont{R.~R.} \bibnamefont{Ruel}},
  \bibinfo{author}{\bibfnamefont{P.~B.} \bibnamefont{Littlewood}},
  \bibinfo{author}{\bibfnamefont{K.~W.} \bibnamefont{West}}, \bibnamefont{and}
  \bibinfo{author}{\bibfnamefont{L.~N.} \bibnamefont{Pfeiffer}},
  \bibinfo{journal}{Phys. Rev. B} \textbf{\bibinfo{volume}{45}},
  \bibinfo{pages}{13784} (\bibinfo{year}{1992}).

\bibitem[{\citenamefont{Ye et~al.}(2002)\citenamefont{Ye, Engel, Tsui, Lewis,
  Pfeiffer, and West}}]{yewc}
\bibinfo{author}{\bibfnamefont{P.~D.} \bibnamefont{Ye}},
  \bibinfo{author}{\bibfnamefont{L.~W.} \bibnamefont{Engel}},
  \bibinfo{author}{\bibfnamefont{D.~C.} \bibnamefont{Tsui}},
  \bibinfo{author}{\bibfnamefont{R.~M.} \bibnamefont{Lewis}},
  \bibinfo{author}{\bibfnamefont{L.~N.} \bibnamefont{Pfeiffer}},
  \bibnamefont{and} \bibinfo{author}{\bibfnamefont{K.~W.} \bibnamefont{West}},
  \bibinfo{journal}{Phys. Rev. Lett.} \textbf{\bibinfo{volume}{89}},
  \bibinfo{pages}{176802} (\bibinfo{year}{2002}).

\bibitem[{\citenamefont{Shayegan}(1997)}]{msreview}
\bibinfo{author}{\bibfnamefont{M.}~\bibnamefont{Shayegan}},
  \emph{\bibinfo{title}{{\rm in }Perspectives in Quantum Hall Effects, {\rm
  edited by S. Das Sarma and A. Pinczuk}}}
  (\bibinfo{publisher}{Wiley-Interscience, New York}, \bibinfo{year}{1997}), p.
  \bibinfo{pages}{343}.

\bibitem[{\citenamefont{Chen et~al.}(2003)\citenamefont{Chen, Lewis, Engel,
  Tsui, Ye, Pfeiffer, and West}}]{chen:2003}
\bibinfo{author}{\bibfnamefont{Y.}~\bibnamefont{Chen}},
  \bibinfo{author}{\bibfnamefont{R.~M.} \bibnamefont{Lewis}},
  \bibinfo{author}{\bibfnamefont{L.~W.} \bibnamefont{Engel}},
  \bibinfo{author}{\bibfnamefont{D.~C.} \bibnamefont{Tsui}},
  \bibinfo{author}{\bibfnamefont{P.~D.} \bibnamefont{Ye}},
  \bibinfo{author}{\bibfnamefont{L.~N.} \bibnamefont{Pfeiffer}},
  \bibnamefont{and} \bibinfo{author}{\bibfnamefont{K.~W.} \bibnamefont{West}},
  \bibinfo{journal}{Phys. Rev. Lett.} \textbf{\bibinfo{volume}{91}},
  \bibinfo{pages}{016801} (\bibinfo{year}{2003}).

\bibitem[{\citenamefont{Hatke et~al.}(2014)\citenamefont{Hatke, Liu, Magill,
  Moon, Engel, Shayegan, Pfeiffer, West, and Baldwin}}]{hatke:2014}
\bibinfo{author}{\bibfnamefont{A.~T.} \bibnamefont{Hatke}},
  \bibinfo{author}{\bibfnamefont{Y.}~\bibnamefont{Liu}},
  \bibinfo{author}{\bibfnamefont{B.~A.} \bibnamefont{Magill}},
  \bibinfo{author}{\bibfnamefont{B.~H.} \bibnamefont{Moon}},
  \bibinfo{author}{\bibfnamefont{L.~W.} \bibnamefont{Engel}},
  \bibinfo{author}{\bibfnamefont{M.}~\bibnamefont{Shayegan}},
  \bibinfo{author}{\bibfnamefont{L.~N.} \bibnamefont{Pfeiffer}},
  \bibinfo{author}{\bibfnamefont{K.~W.} \bibnamefont{West}}, \bibnamefont{and}
  \bibinfo{author}{\bibfnamefont{K.~W.} \bibnamefont{Baldwin}},
  \bibinfo{journal}{Nat. Commun.} \textbf{\bibinfo{volume}{5}},
  \bibinfo{pages}{4154} (\bibinfo{year}{2014}).

\bibitem[{\citenamefont{Tiemann et~al.}(2014)\citenamefont{Tiemann, Rhone,
  Shibata, and Muraki}}]{tiemann:2014}
\bibinfo{author}{\bibfnamefont{L.}~\bibnamefont{Tiemann}},
  \bibinfo{author}{\bibfnamefont{T.~D.} \bibnamefont{Rhone}},
  \bibinfo{author}{\bibfnamefont{N.}~\bibnamefont{Shibata}}, \bibnamefont{and}
  \bibinfo{author}{\bibfnamefont{K.}~\bibnamefont{Muraki}},
  \bibinfo{journal}{Nat. Phys.} \textbf{\bibinfo{volume}{10}},
  \bibinfo{pages}{648} (\bibinfo{year}{2014}).

\bibitem[{\citenamefont{Zhu et~al.}(2010)\citenamefont{Zhu, Chen, Jiang, Engel,
  Tsui, Pfeiffer, and West}}]{zhu:2010}
\bibinfo{author}{\bibfnamefont{H.}~\bibnamefont{Zhu}},
  \bibinfo{author}{\bibfnamefont{Y.~P.} \bibnamefont{Chen}},
  \bibinfo{author}{\bibfnamefont{P.}~\bibnamefont{Jiang}},
  \bibinfo{author}{\bibfnamefont{L.~W.} \bibnamefont{Engel}},
  \bibinfo{author}{\bibfnamefont{D.~C.} \bibnamefont{Tsui}},
  \bibinfo{author}{\bibfnamefont{L.~N.} \bibnamefont{Pfeiffer}},
  \bibnamefont{and} \bibinfo{author}{\bibfnamefont{K.~W.} \bibnamefont{West}},
  \bibinfo{journal}{Phys. Rev. Lett.} \textbf{\bibinfo{volume}{105}},
  \bibinfo{pages}{126803} (\bibinfo{year}{2010}).

\bibitem[{\citenamefont{Manoharan et~al.}(1996)\citenamefont{Manoharan, Suen,
  Santos, and Shayegan}}]{manoharan:1996}
\bibinfo{author}{\bibfnamefont{H.~C.} \bibnamefont{Manoharan}},
  \bibinfo{author}{\bibfnamefont{Y.~W.} \bibnamefont{Suen}},
  \bibinfo{author}{\bibfnamefont{M.~B.} \bibnamefont{Santos}},
  \bibnamefont{and} \bibinfo{author}{\bibfnamefont{M.}~\bibnamefont{Shayegan}},
  \bibinfo{journal}{Phys. Rev. Lett.} \textbf{\bibinfo{volume}{77}},
  \bibinfo{pages}{1813} (\bibinfo{year}{1996}).

\bibitem[{\citenamefont{Thiebaut et~al.}(2015)\citenamefont{Thiebaut, Regnault,
  and Goerbig}}]{goerbigwqw}
\bibinfo{author}{\bibfnamefont{N.}~\bibnamefont{Thiebaut}},
  \bibinfo{author}{\bibfnamefont{N.}~\bibnamefont{Regnault}}, \bibnamefont{and}
  \bibinfo{author}{\bibfnamefont{M.~O.} \bibnamefont{Goerbig}},
  \bibinfo{journal}{Phys. Rev. B} \textbf{\bibinfo{volume}{92}},
  \bibinfo{pages}{245401} (\bibinfo{year}{2015}).

\bibitem[{\citenamefont{Chen et~al.}(2004)\citenamefont{Chen, Lewis, Engel,
  Tsui, Ye, Wang, Pfeiffer, and West}}]{chen:2004}
\bibinfo{author}{\bibfnamefont{Y.~P.} \bibnamefont{Chen}},
  \bibinfo{author}{\bibfnamefont{R.~M.} \bibnamefont{Lewis}},
  \bibinfo{author}{\bibfnamefont{L.~W.} \bibnamefont{Engel}},
  \bibinfo{author}{\bibfnamefont{D.~C.} \bibnamefont{Tsui}},
  \bibinfo{author}{\bibfnamefont{P.~D.} \bibnamefont{Ye}},
  \bibinfo{author}{\bibfnamefont{Z.~H.} \bibnamefont{Wang}},
  \bibinfo{author}{\bibfnamefont{L.~N.} \bibnamefont{Pfeiffer}},
  \bibnamefont{and} \bibinfo{author}{\bibfnamefont{K.~W.} \bibnamefont{West}},
  \bibinfo{journal}{Phys. Rev. Lett.} \textbf{\bibinfo{volume}{93}},
  \bibinfo{pages}{206805} (\bibinfo{year}{2004}).

\bibitem[{\citenamefont{Hatke et~al.}(2015)\citenamefont{Hatke, Liu, Engel,
  Shayegan, Pfeiffer, West, and Baldwin}}]{hatke:2015}
\bibinfo{author}{\bibfnamefont{A.~T.} \bibnamefont{Hatke}},
  \bibinfo{author}{\bibfnamefont{Y.}~\bibnamefont{Liu}},
  \bibinfo{author}{\bibfnamefont{L.~W.} \bibnamefont{Engel}},
  \bibinfo{author}{\bibfnamefont{M.}~\bibnamefont{Shayegan}},
  \bibinfo{author}{\bibfnamefont{L.~N.} \bibnamefont{Pfeiffer}},
  \bibinfo{author}{\bibfnamefont{K.~W.} \bibnamefont{West}}, \bibnamefont{and}
  \bibinfo{author}{\bibfnamefont{K.~W.} \bibnamefont{Baldwin}},
  \bibinfo{journal}{Nat. Commun.} \textbf{\bibinfo{volume}{6}},
  \bibinfo{pages}{7071} (\bibinfo{year}{2015}).

\bibitem[{\citenamefont{Fukuyama and Lee}(1978)}]{fukuyama:1978}
\bibinfo{author}{\bibfnamefont{H.}~\bibnamefont{Fukuyama}} \bibnamefont{and}
  \bibinfo{author}{\bibfnamefont{P.~A.} \bibnamefont{Lee}},
  \bibinfo{journal}{Phys. Rev. B} \textbf{\bibinfo{volume}{18}},
  \bibinfo{pages}{6245} (\bibinfo{year}{1978}).

\bibitem[{\citenamefont{Chitra et~al.}(2001)\citenamefont{Chitra, Giamarchi,
  and Doussal}}]{chitra:2001}
\bibinfo{author}{\bibfnamefont{R.}~\bibnamefont{Chitra}},
  \bibinfo{author}{\bibfnamefont{T.}~\bibnamefont{Giamarchi}},
  \bibnamefont{and} \bibinfo{author}{\bibfnamefont{P.~L.}
  \bibnamefont{Doussal}}, \bibinfo{journal}{Phys. Rev. B}
  \textbf{\bibinfo{volume}{65}}, \bibinfo{pages}{035312}
  (\bibinfo{year}{2001}).

\bibitem[{\citenamefont{Fertig}(1999)}]{fertig:1999}
\bibinfo{author}{\bibfnamefont{H.~A.} \bibnamefont{Fertig}},
  \bibinfo{journal}{Phys. Rev. B} \textbf{\bibinfo{volume}{59}},
  \bibinfo{pages}{2120} (\bibinfo{year}{1999}).

\bibitem[{\citenamefont{Fogler and Huse}(2000)}]{fogler:2000}
\bibinfo{author}{\bibfnamefont{M.~M.} \bibnamefont{Fogler}} \bibnamefont{and}
  \bibinfo{author}{\bibfnamefont{D.~A.} \bibnamefont{Huse}},
  \bibinfo{journal}{Phys. Rev. B} \textbf{\bibinfo{volume}{62}},
  \bibinfo{pages}{7553} (\bibinfo{year}{2000}).

\end{thebibliography}

\end{document}